\documentclass[aps,prl,preprint,superscriptaddress]{revtex4-2}
\usepackage[pdftex]{graphicx}
\DeclareGraphicsExtensions{.pdf,.jpeg,.png}
\usepackage{epstopdf}
\usepackage[colorlinks=true,linkcolor=blue,citecolor=blue]{hyperref}

\begin{document}

\title{Interfacial-hybridization-modified Ir Ferromagnetism and Electronic Structure in LaMnO$_3$/SrIrO$_3$ Superlattices}

\author{Yujun~Zhang}
\email{zhangyujun@ihep.ac.cn}
\affiliation{Institute of High Energy Physics, Chinese Academy of Sciences, Yuquan Road 19B, Shijingshan District, Beijing, 100049, China}
\affiliation{Graduate School of Materials Science, University of Hyogo, 3-2-1 Kouto, Kamigori-cho, Ako-gun, Hyogo 678-1297, Japan}
\affiliation{Institute for Solid State Physics, University of Tokyo, 5-1-5 Kashiwanoha, Chiba 277-8581, Japan}
\author{Yong~Zheng~Luo}
\email{mpeluoy@nus.edu.sg}
\affiliation{Department of Mechanical Engineering, National University of Singapore, 117575, Singapore}
\author{Liang~Wu}
\affiliation{State Key Lab of New Ceramics and Fine Processing, School of Materials Science and Engineering, Tsinghua University, Beijing 100084, China}
\affiliation{Department of Physics and Astronomy, Rutgers University, Piscataway, NJ 08854, USA}
\author{Motohiro Suzuki}
\affiliation{JASRI, 1-1-1 Kouto, Sayo, Hyogo 679-5198, Japan}
\author{Yasuyuki~Hirata}
\affiliation{Institute for Solid State Physics, University of Tokyo, 5-1-5 Kashiwanoha, Chiba 277-8581, Japan}
\author{Kohei~Yamagami}
\affiliation{Institute for Solid State Physics, University of Tokyo, 5-1-5 Kashiwanoha, Chiba 277-8581, Japan}
\author{Kou~Takubo}
\affiliation{Institute for Solid State Physics, University of Tokyo, 5-1-5 Kashiwanoha, Chiba 277-8581, Japan}
\author{Keisuke~Ikeda}
\affiliation{Institute for Solid State Physics, University of Tokyo, 5-1-5 Kashiwanoha, Chiba 277-8581, Japan}
\author{Kohei~Yamamoto}
\affiliation{Institute for Solid State Physics, University of Tokyo, 5-1-5 Kashiwanoha, Chiba 277-8581, Japan}
\author{Akira~Yasui}
\affiliation{JASRI, 1-1-1 Kouto, Sayo, Hyogo 679-5198, Japan}
\author{Naomi~Kawamura}
\affiliation{JASRI, 1-1-1 Kouto, Sayo, Hyogo 679-5198, Japan}
\author{Chun~Lin}
\affiliation{Department of Physics, University of Tokyo, Bunkyo-ku, Tokyo 113-0033, Japan}
\author{Keisuke Koshiishi}
\affiliation{Department of Physics, University of Tokyo, Bunkyo-ku, Tokyo 113-0033, Japan}
\author{Xin~Liu}
\affiliation{Department of Physics, Beijing Normal University, Beijing 100875, China}
\author{Jinxing~Zhang}
\affiliation{Department of Physics, Beijing Normal University, Beijing 100875, China}
\author{Yasushi~Hotta}
\affiliation{Department of Engineering, University of Hyogo, 2167 Shosha, Himeji, Hyogo 671-2280, Japan}
\author{X.~Renshaw~Wang}
\affiliation{School of Physical and Mathematical Sciences \& School of Electrical and Electronic Engineering, Nanyang Technological University, Singapore 637371, Singapore}
\author{Atsushi~Fujimori}
\affiliation{Department of Physics, University of Tokyo, Bunkyo-ku, Tokyo 113-0033, Japan}
\author{Yuanhua~Lin}
\affiliation{State Key Lab of New Ceramics and Fine Processing, School of Materials Science and Engineering, Tsinghua University, Beijing 100084, China}
\author{Cewen~Nan}
\affiliation{State Key Lab of New Ceramics and Fine Processing, School of Materials Science and Engineering, Tsinghua University, Beijing 100084, China}
\author{Lei~Shen}
\affiliation{Department of Mechanical Engineering, National University of Singapore, 117575, Singapore}
\author{Hiroki~Wadati}
\affiliation{Graduate School of Materials Science, University of Hyogo, 3-2-1 Kouto, Kamigori-cho, Ako-gun, Hyogo 678-1297, Japan}
\affiliation{Institute for Solid State Physics, University of Tokyo, 5-1-5 Kashiwanoha, Chiba 277-8581, Japan}

\date{\today}

\begin{abstract}
Artificially fabricated 3$d$/5$d$ superlattices (SLs) involve both strong electron correlation and spin-orbit coupling in one material by means of interfacial 3$d$-5$d$ coupling, whose mechanism remains mostly unexplored. In this work we investigated the mechanism of interfacial coupling in LaMnO$_3$/SrIrO$_3$ SLs by several spectroscopic approaches. Hard x-ray absorption, magnetic circular dichroism and photoemission spectra evidence the systematic change of the Ir ferromagnetism and the electronic structure with the change of the SL repetition period. First-principles calculations further reveal the mechanism of the SL-period dependence of the interfacial electronic structure and the local properties of the Ir moments, confirming that the formation of Ir-Mn molecular orbital is responsible for the interfacial coupling effects. The SL-period dependence of the ratio between spin and orbital components of the Ir magnetic moments can be attributed to the realignment of electron spin during the formation of the interfacial molecular orbital. Our results clarify the nature of interfacial coupling in this prototypical 3$d$/5$d$ SL system and the conclusion will shed light on the study of other strongly correlated and spin-orbit coupled oxide hetero-interfaces.
\end{abstract}

\maketitle

\section*{Introduction}Entanglement of charge, spin, lattice and orbital degrees of freedom in transition metal oxides (TMOs) has attracted a great amount of research attentions recently~\cite{1_Imada,2_Dagotto,3_Quintanilla,4_Keimer,5_Ramirez}. 
Strong electron correlation in TMOs is a necessity to support the existence of local magnetic moments and various magnetic/charge/orbital orderings. Meanwhile, novel topics related to spin-orbit coupling (SOC) have also become research hot spots in condensed matter physics~\cite{6_Soumyanarayanan,7_Qi}. SOC can work in real space and give rise to a variety of non-collinear magnetic structures such as skyrmions~\cite{6_Soumyanarayanan,8_Muhlbauer}, or in reciprocal $k$-space to produce topologically non-trivial band structures~\cite{6_Soumyanarayanan,7_Qi,9_Bansil}. Consequently, fabrication and investigation of systems involving both of these two interactions become of not only scientific but also technical interest~\cite{10_Pesin,11_Kim}. Local magnetic moments and orderings within a strong SOC regime still remain mostly unexplored. TMOs with heavy transition metals, such as iridates, are ideal hosts for coexistence of both significant electron correlation and SOC. An unprecedented $J$\textsubscript{eff}=1/2 Mott insulating state with canted antiferromagnetic (AFM) ordering was realized in layered perovskite iridates such as Sr$_2$IrO$_4$~\cite{11_Kim,12_Moon,13_Kim} and Sr$_3$Ir$_2$O$_7$~\cite{12_Moon,14_Fujiyama}, where the collaboration of strong electron correlation and SOC plays a crucial role. Nevertheless, generally 5$d$ TMOs are not capable to sustain magnetic orderings due to the large spatial extension of the 5$d$ electrons~\cite{15_Cao}. While their well-investigated 3$d$ TMO counterparts usually possess weak SOC, even though the strength of electron correlation is always sufficient to support magnetism.

Interfaces between dissimilar materials can provide an intriguing playground for manipulation of various physical properties~\cite{16_Hwang,17_Zubko,18_Matsuno,19_Ohmoto,20_Hellman}. Great improvement of thin-film fabrication techniques enables accurately controlled design of epitaxial TMO heterostructures and SLs with atomically abrupt interfaces. It appears to be a natural strategy that artificial 3$d$/5$d$ TMO heterostructures or SLs are promising candidates to involve both significant electron correlation and SOC simultaneously. Pioneering research about 3$d$/5$d$ SLs was triggered by investigation on SrIrO$_3$/SrTiO$_3$ (SIO/STO) perovskite SLs~\cite{21_Matsuno}, as a comparison with the Ruddlesden-Popper series iridates Sr$_{n+1}$Ir$_n$O$_{3n+1}$. 
Meanwhile, strong interfacial 3$d$-5$d$ coupling was reported in La$_{1-x}$Sr$_{x}$MnO$_3$/SrIrO$_3$ (0$<$\textit{x}$<$1, LSMO/SIO) SLs~\cite{22_Nichols,23_Yi,24_Yi,25_Kim,26_Huang}. Emergent Ir ferromagnetic (FM) moments can be induced by the interfacial coupling with Mn FM moments, and in turn the magnetic properties of Mn layers can be significantly modified as well. Perpendicular magnetic anisotropy and concomitant anomalous Hall effect were observed in \mbox{$x=1$} SLs~\cite{22_Nichols}, and modulation of magnetic anisotropy in LSMO layers was also studied~\cite{23_Yi,24_Yi}. Recent reports claim that interfacial hybridization between Ir and Mn is responsible for the charge transfer in $x=1$ SLs~\cite{27_Okamoto} and spectroscopic properties of $x=0.33$ SLs~\cite{25_Kim}.

Perovskite SLs are ideal systems for investigation of interfacial coupling mechanisms thanks to their high interface quality. The modification of electronic structure at the interfaces can lead to consequent change of magnetic properties. 
Research on the SL-period dependent evolution of the interfacial electronic structure will be informative for understanding the role of interfacial 3$d$-5$d$ coupling to affect the Ir magnetism, which has not been systematically investigated so far. For this purpose, we fabricated LaMnO$_3$/SrIrO$_3$ (LMO/SIO) SLs with different repetition periods. X-ray magnetic circular dichroism (XMCD) were employed to study how the SL period and the interfacial coupling can affect the properties of FM Ir moments. X-ray absorption spectra (XAS) and hard x-ray photoemission spectroscopy (HAXPES) were carried out to characterize the SL-period dependence of the electronic structure. A systematic SL-period dependent trend of the ratio between orbital and spin magnetic moments of Ir as well as the electronic structure were observed. First-principles calculations demonstrate a satisfactory consistency with the experimental results and reveal that the formation of the interfacial Ir-Mn molecular orbital associated with concomitant electronic-structure change is the pivotal mechanism behind this interfacial coupling.

\section*{Methods}
[(LMO)$_a$/(SIO)$_a$]$_b$ (($a$,$b$)=(1,24), (2,12) and (8,3), where $a$ is counted in unit cells, SL$aa$ in abbreviation) SL samples as well as LMO and SIO reference samples (24 unit cells) were fabricated by laser molecular beam epitaxy. A KrF excimer pulsed laser ($\lambda$=248~nm) with a repetition rate of 2~Hz and an energy density of $\sim$1.5~J/cm$^2$ was employed. The sample temperature and ambient oxygen pressure were controlled at 720~\textsuperscript{o}C and 16~Pa, respectively. The distance between the stoichiometric LMO or SIO targets and STO(001) single crystal substrates was set at 6 cm. The layer-by-layer growth of the SLs was guaranteed by monitoring the oscillation of the reflection high-energy electron diffraction signal (as reported in Ref.~\cite{26_Huang}). The sample structure is schematically displayed in Fig.~\ref{fig1}(a). The crystal structure of the SLs was characterized by an x-ray diffractometer with Cu $K_{\alpha}$ radiation~(XRD, Rigaku RINT-2200). Basic magnetic properties of the SLs were characterized by a superconducting quantum interference device (SQUID, Quantum Design).

The Ir $L$ edge XAS/XMCD measurements were conducted at BL39XU of SPring-8. A He-flowing cryostat was used to cool the samples to a lowest temperature of 30~K. In-plane magnetic field up to 2~T along the x-ray propagation was applied by an electromagnet. The Ir $L_{3,2}$ edge XAS/XMCD spectra were collected by standard helicity reversal technique~\cite{28_Suzuki} with a grazing incidence geometry (5.5\textsuperscript{o} incidence angle) and partial fluorescence yield (PFY) mode. For PFY detection of XAS/XMCD at the Ir $L_3$ and $L_2$ edges, Ir $L_{\alpha}$ and $L_{\beta}$ emissions were collected and energy-analyzed respectively by a four-element silicon drift detector (Sirius 4, SGX Sensortech Inc.). 
The positive magnetic field direction is defined opposite to the x-ray propagation. HAXPES measurements were carried out at BL47XU of SPring-8. The incidence angle of 7.94~keV hard x-ray was set at $\sim$1\textsuperscript{o} and the emitted photoelectrons were detected by a Scienta R-4000 electron energy analyzer, whose energy resolution was $\sim$280~meV.

First-principles calculations were carried out within the framework of density functional theory (DFT) \cite{29_Hohenberg,30_Kohn} using the generalized gradient approximation (GGA) in the parameterization of Perdew-Burke-Ernzerhof (PBE) format exchange-correlation functional \cite{31_Perdew}, as implemented in the Vienna $ab$ $initio$ Simulation Package (VASP) \cite{32_Kresse,33_Kresse,34_Kresse,35_Kresse}.  SOC is implemented in the projector augmented wave (PAW) method \cite{36_Blochl,37_Kresse} which is based on a transformation that maps all electron wave functions to smooth pseudowave functions to describe the interaction between electrons and ions. The corresponding electronic configurations for each element are Sr: 4$s$4$p$5$s$; Ir: 5$d$6$s$; O: 2$s$2$p$; La: 5$s$5$p$6$s$5$d$; Mn: 3$p$3$d$4$s$. The cutoff energy is set to 500~eV. To account for strong correlation effects \cite{38_Liechtenstein},  we included the Hubbard correction $U$ for Ir and Mn $d$ states with $U$\textsubscript{Ir$_{5d}$}=2~eV and $U$\textsubscript{Mn$_{3d}$}=3~eV~\cite{39_Caviglia,27_Okamoto}.  We used $4\times4\times4$ K-points following the Monkhorst-Pack scheme in our systems. The convergence criterion for the electronic relaxation is $10^{-6}$~eV. In this calculation, we relaxed the SL cell parameters and atomic positions with the in-plane lattice constant constrained to that of STO. 
The doubled unit cell has been used with the experimental in-plane lattice constant of STO, $a$=$b$=3.905$\times\sqrt{2}$~\AA~(see Fig.~\ref{fig7} in the Appendix A).
Optimized SL structures were achieved when forces on all the atoms were $<$0.01 eV/\AA.  
We calculated SL11, SL22, SL33 and SL44 rather than SL11, SL22 and SL88 investigated in our experiments since the supercell of SL88 is too large for DFT-based calculations. 

\section*{Results and Discussions}
\begin{figure}
	\includegraphics[width=\columnwidth]{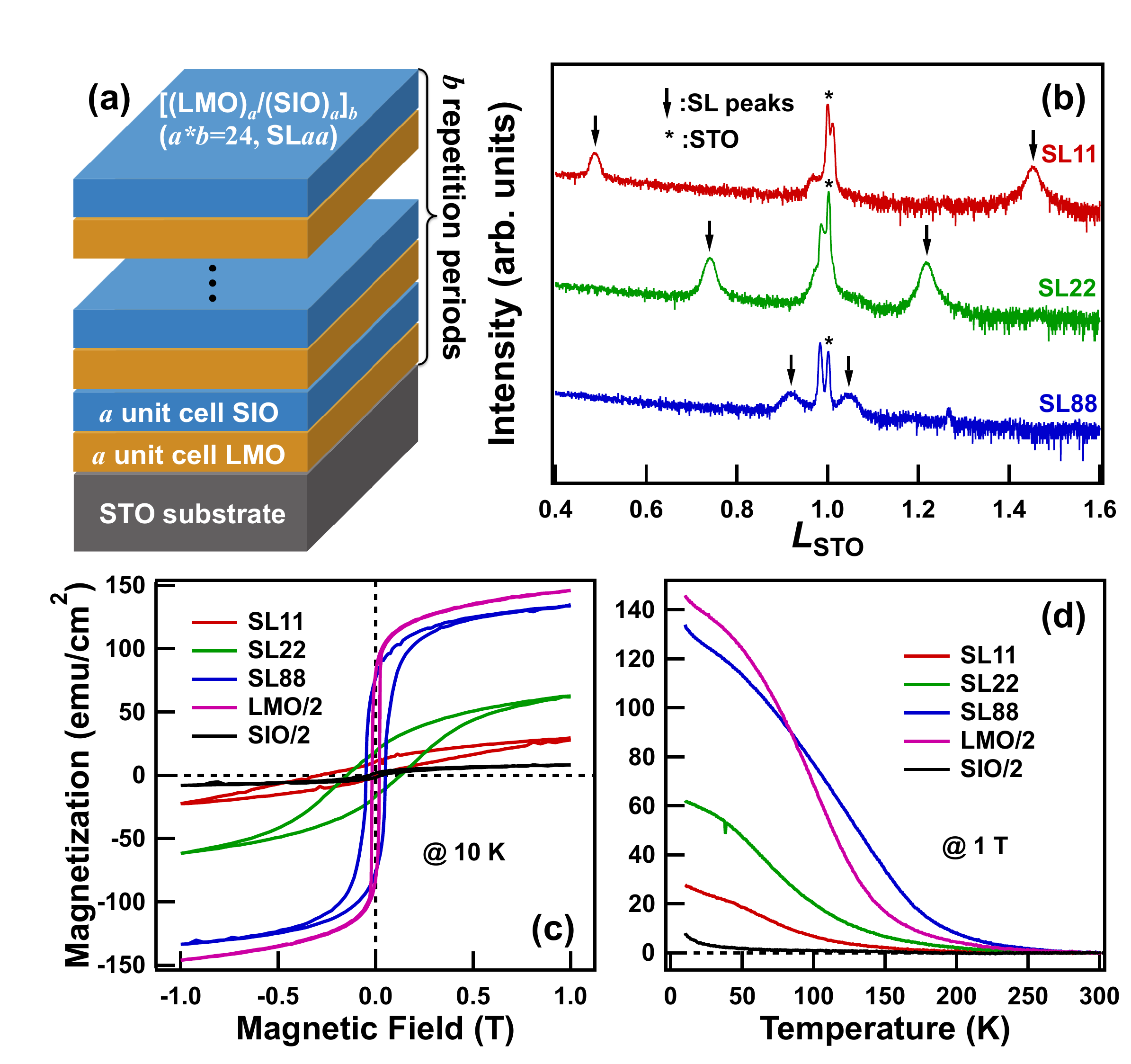}
	\caption{(a) Structure schematic of the LMO/SIO SLs. (b) XRD (0~0~$L$) scans of the SLs. SL satellite peaks and (0~0~1) peaks of STO substrate are indicated. The peaks near the substrate diffraction originate from the fundamental (0~0~1) diffractions of LMO and SIO. (c) $M$-$H$ and (d) $M$-$T$ curves of the SLs and the reference samples. The magnetic field was applied in the in-plane [001] direction. The magnetization of reference samples is divided by a factor of 2 for comparison, so that each curve shows the magnetization which includes the same amount of Ir or Mn.}
	\label{fig1}
\end{figure}
The XRD $L$ scan spectra in (0~0~$L$) direction of the SLs are presented in Fig.~\ref{fig1}(b). SL satellite peaks can be clearly observed beside the (0~0~1) diffraction of the STO substrate, confirming the high quality of the SLs. In-plane magnetization-field ($M$-$H$) and magnetization-temperature ($M$-$T$) curves shown in Fig.~\ref{fig1}(c,d) indicate FM behaviors of all the three SLs, as well as the pure LMO reference sample. The FM Curie temperatures ($T_c$) are around 150~K, 170~K and 200~K for SL11, SL22 and SL88, respectively. The saturated magnetization and $T_c$ of SLs change systematically with the SL period, which should be attributed to the enhancement of Mn-Mn electron hopping within LMO layers as the SL period increases. The reference SIO sample remains paramagnetic down to 10~K as previously reported~\cite{15_Cao}. 
\begin{figure}
	\includegraphics[width=\columnwidth]{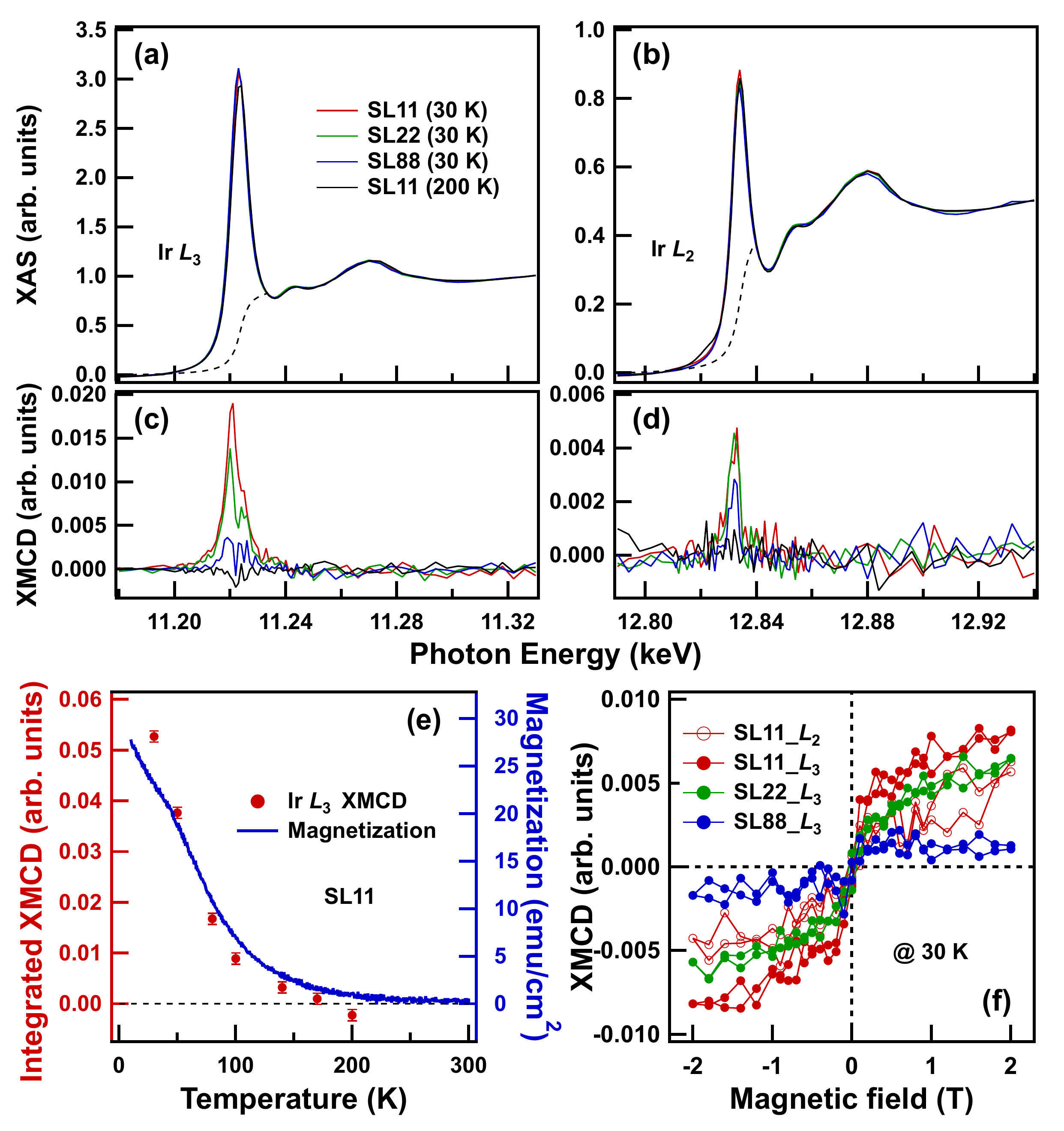}
	\caption{(a,b) XAS and (c,d) XMCD results of LMO/SIO SLs at the Ir $L$ edge with 2~T magnetic field applied. The area between XAS and the arctangent background (dashed lines) is estimated as the white-line intensity. (e) Temperature dependence of the integrated Ir $L_3$ edge XMCD intensity  of SL11 compared with its $M$-$T$ curve. (f) Element-specific magnetization curves of Ir with the photon energy set at the XMCD peaks.}
	\label{fig2}
\end{figure}

To detect and comprehensively analyze the SL-period dependence of the Ir magnetism, hard x-ray XAS/XMCD measurements at the Ir $L_{3,2}$ edges were conducted and the main results are displayed in Fig.~\ref{fig2}. The XAS (defined as $(\mu^{+}+\mu^{-})/2$ and averaged for positive and negative magnetic fields, where $\mu^+$ and $\mu^-$ are PFY-XAS measured by x-rays with left and right helicities) in Fig.~\ref{fig2}(a,b) of all the SLs have similar line shape with white-line peaks at around 11.223~keV($L_3$) and 12.834~keV($L_2$), indicating a localized feature of Ir 5$d$ states. No obvious peak shift induced by interfacial charge transfer between Ir and Mn was observed, which is consistent with previous reports~\cite{40_Zhong}. Ir-Mn charge transfer will lead to deviation of the Ir valence from the nominal valence state of 4+
and add complexity to the investigation. So we chose pure LMO to construct the SLs since it is reported that with the increase of the Sr\%, significant charge transfer between Ir and Mn will appear in LSMO/SIO SL system~\cite{22_Nichols,27_Okamoto}.  The XAS peak intensity slightly changes in different SLs, which suggests possible modification of the spin-orbital states of Ir and will be detailedly discussed in the latter parts. 

The XMCD signal in Fig.~\ref{fig2}(c,d) is defined as $\mu^{+}-\mu^{-}$ (averaged for positive and negative magnetic fields). All three SLs exhibit clear XMCD signal at 30 K. The XMCD peak positions are located at $\sim$2~eV lower than the XAS white-line peaks (about 11.221~keV for $L_3$ and 12.832~keV for $L_2$), which is a sign of intrinsic XMCD signal mainly originating from $t_{2g}$ states at lower energy, rather than artifacts from the XAS measurement. The positive XMCD signals at both the $L_3$ and  $L_2$ edges and the larger intensity of the $L_3$ XMCD peak indicate that the net magnetization of Ir is antiparallel to the external field and antiferromagnetically coupled to the Mn magnetization as previously reported~\cite{22_Nichols,23_Yi,24_Yi,25_Kim}. The XAS/XMCD spectra of SL11 measured at temperature above the FM transition (200~K) are also displayed in Fig.~\ref{fig2}. XMCD signal is negligible at both the $L_3$ and $L_2$ edges, evidencing the disappearance of FM ordering of Ir moments. 

Temperature dependent XMCD measurements of SL11 at the Ir $L_3$ edge (Fig.~\ref{fig2}(e)) shows that the integrated XMCD intensity decreases with the increase of temperature and vanishes at $\sim$150~K, in consistency with the $T_c$ of the SL11 sample. The $M$-$T$ curve of SL11 is also plotted into Fig.~\ref{fig2}(e) for comparison. Since the total magnetization is dominated by the Mn moments, this result confirms that the emergent FM Ir moments originate from the interfacial coupling and rely on the existence of Mn FM ordering. Fig.~\ref{fig2}(f) shows the magnetic field dependence of Ir XMCD signal. Clear FM behaviors were observed for all the SLs.
\begin{figure}
	\includegraphics[width=\columnwidth]{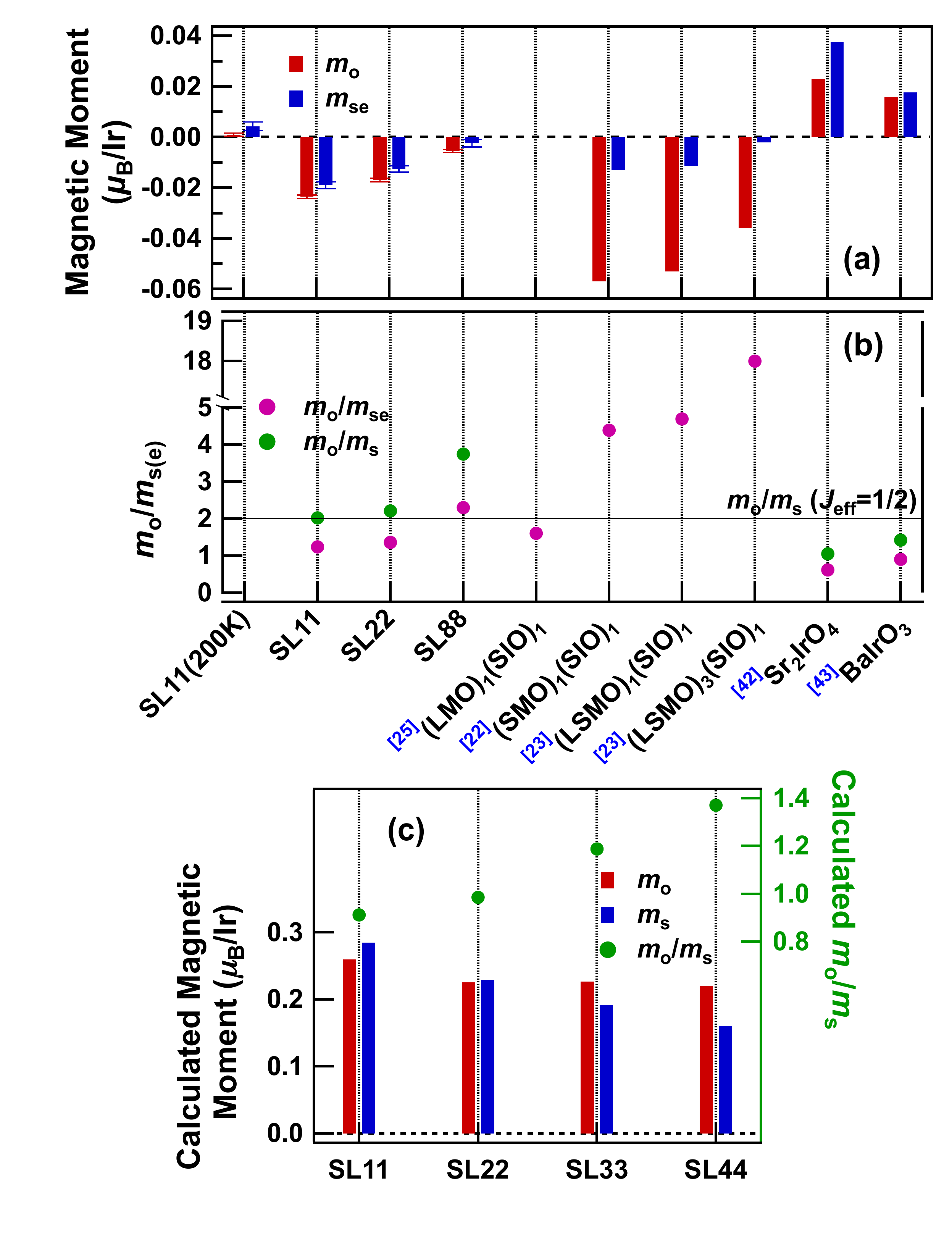}
	\caption{(a) $m_o$ and $m_{se}$ of Ir obtained by sum-rules analysis of the XMCD results. (b) $m_o$/$m_{s(e)}$ ratios of the SLs. $m_o$/$m_s$ ratios are estimated by $\left<T_z\right>/\left<S_z\right>$=0.18~\cite{43_Haskel,44_LagunaMarco}. The results are compared with previously reported sum-rules analyses of Ir magnetism~\cite{22_Nichols,23_Yi,25_Kim,43_Haskel,44_LagunaMarco}. (c) $m_o$, $m_s$ and $m_o$/$m_s$ ratios calculated by DFT.}
	\label{fig3}
\end{figure}

Based on the successful observation of FM XMCD signal at the Ir $L_{3,2}$ edges, sum-rules analysis can provide important information about the local properties of the Ir magnetic moments. By applying sum-rules analysis with the following formulas~\cite{41_Thole,42_Carra}, it can be obtained that at 30~K the orbital magnetic moments of Ir ($m_o$) are -0.0235, -0.0170 and -0.0055 $\mu$\textsubscript{B}/atom, and the effective spin magnetic moments of Ir ($m_{se}$) are -0.0190, -0.0125 and -0.0024 $\mu$\textsubscript{B}/atom, for SL11, SL22 and SL88 respectively. With the assumption of the negligible charge transfer between Ir and Mn according to the previous theoretical report~\cite{40_Zhong}, Ir has a nominal valence state of 4+ in our SL system, so the number of 5$d$ holes is estimated as $n_{h}=5$. 

$$	m_o=-\frac{4\int_{(L_{2}+L_{3})} (\mu^+-\mu^-) \,dE}{3\int_{(L_{2}+L_{3})} (\mu^++\mu^-) \,dE}n_h\;(\mu_B)$$ 
$$m_{se}=-\frac{2\int_{L_{3}} (\mu^+-\mu^-) \,dE-4\int_{L_{2}} (\mu^+-\mu^-) \,dE}{\int_{(L_{2}+L_{3})} (\mu^++\mu^-) \,dE}n_h\;(\mu_B)
$$

Fig.~\ref{fig3}(a) displays the sum-rules analysis results of our samples as well as previous reports of other LSMO/SIO SL and magnetic iridate systems~\cite{22_Nichols,23_Yi,25_Kim,43_Haskel,44_LagunaMarco}. 
The size of measured Ir FM moments is quite consistent with the previous reports~\cite{22_Nichols,23_Yi,24_Yi,25_Kim} that Mn moments dominate the total magnetization and Ir magnetization is 1 or 2 orders of magnitude smaller than Mn. Since FM Ir moments are antiparallel to the external field, the signs of the Ir magnetic moments are negative, which is opposite to the perovskite iridates. 
The size of the Ir moments decreases with the SL period, which suggests that the FM Ir moments mainly distribute near the interfaces (see more detailed comments on this point in the Appendix B). 

Since Ir moments often exhibit canted AFM ordering in perovskite iridates~\cite{45_Kim,21_Matsuno,25_Kim}, the net Ir moment size depends on both the absolute size of the local Ir moment and the canting angle between moments in different AFM sublattices. Hence the net moment size evaluated by XMCD sum-rules analysis varies in different systems (Fig.~\ref{fig3}(a)). While the $m_o$/$m_{se}$ ratio can be compared among different systems (Fig.~\ref{fig3}(b)) and reflects the local properties of the Ir moments. 
First it can be noticed that $m_o$/$m_{se}$ ratio of SL11 is quite consistent with a previous report of the same SL~\cite{24_Yi}, indicating a satisfactory reproducibility of the XMCD measurements. 
Moreover, the $m_o$/$m_{se}$ ratios of LSMO/SIO SL systems are generally much larger than magnetic iridates such as Sr$_2$IrO$_4$ and BaIrO$_3$. This should be attributed to the different origin of Ir moments. The FM Ir moments mainly originate from the interfacial coupling with Mn moments in LSMO/SIO SLs while in magnetic iridates they mainly originate from the electron correlation of the $J$\textsubscript{eff}=1/2 band~\cite{11_Kim}. 

The $m_o$/$m_s$ ratio of ideal $J$\textsubscript{eff}=1/2 model is 2~\cite{11_Kim}, as indicated in Fig.~\ref{fig3}(b). The contributions of both $m_s$ and magnetic dipole term $\left<T_z\right>$ are included in the $m_{se}$ ($m_{se}=m_s+7\left<T_z\right>$), but they can not be easily separated by experiment. To estimate the $m_o$/$m_s$ ratio of our samples and compare with the ideal $J$\textsubscript{eff}=1/2 model, we used the estimation of $\left<T_z\right>/\left<S_z\right>$=0.18 (similar to the values obtained by cluster-model calculations for Sr$_2$IrO$_4$~\cite{43_Haskel} and BaIrO$_3$~\cite{44_LagunaMarco}, $\left<S_z\right>$ is the spin angular momentum) to evaluate the $\left<T_z\right>$ term. 
The estimated $m_o$/$m_s$ ratio of our samples are roughly consistent with the $J$\textsubscript{eff}=1/2 scheme. 

Interfacial Ir-Mn coupling should significantly affect the local properties of the Ir moments, which can be evidenced by the SL-period dependence of the $m_o$/$m_{s(e)}$ ratio of Ir. The $m_o$/$m_{s(e)}$ ratio systematically decreases with decreasing the SL period. The decrease of the SL period leads to the enhancement of the interfacial coupling effects since the volume ratio of interfacial layers increases with the decrease of the SL period. In other words, the $m_o$/$m_{s(e)}$ ratio can be effectively decreased by the interfacial coupling. Remarkably, our DFT calculation results~(Fig.~\ref{fig3}(c)) also show a similar trend that $m_o$/$m_{s}$ ratio systematically decrease with the decreasing of the SL period, which is in good agreement with the experiments. Although the calculated absolute values of $m_o$/$m_s$ ratio deviates from the experimental values, the consistent trend indicates that the local properties of Ir magnetic moments indeed vary systematically with the SL period. It is worth mentioning that the calculated $m_s$ is more strongly dependent on the SL period than $m_o$. The strong SL-period dependence of $m_s$ should be attributed to the orbital reconstruction at the interface, which can make the originally antiparallel electron spins of Ir align parallel. This point will be discussed in detail in the following parts.

\begin{figure}
	\includegraphics[width=\columnwidth]{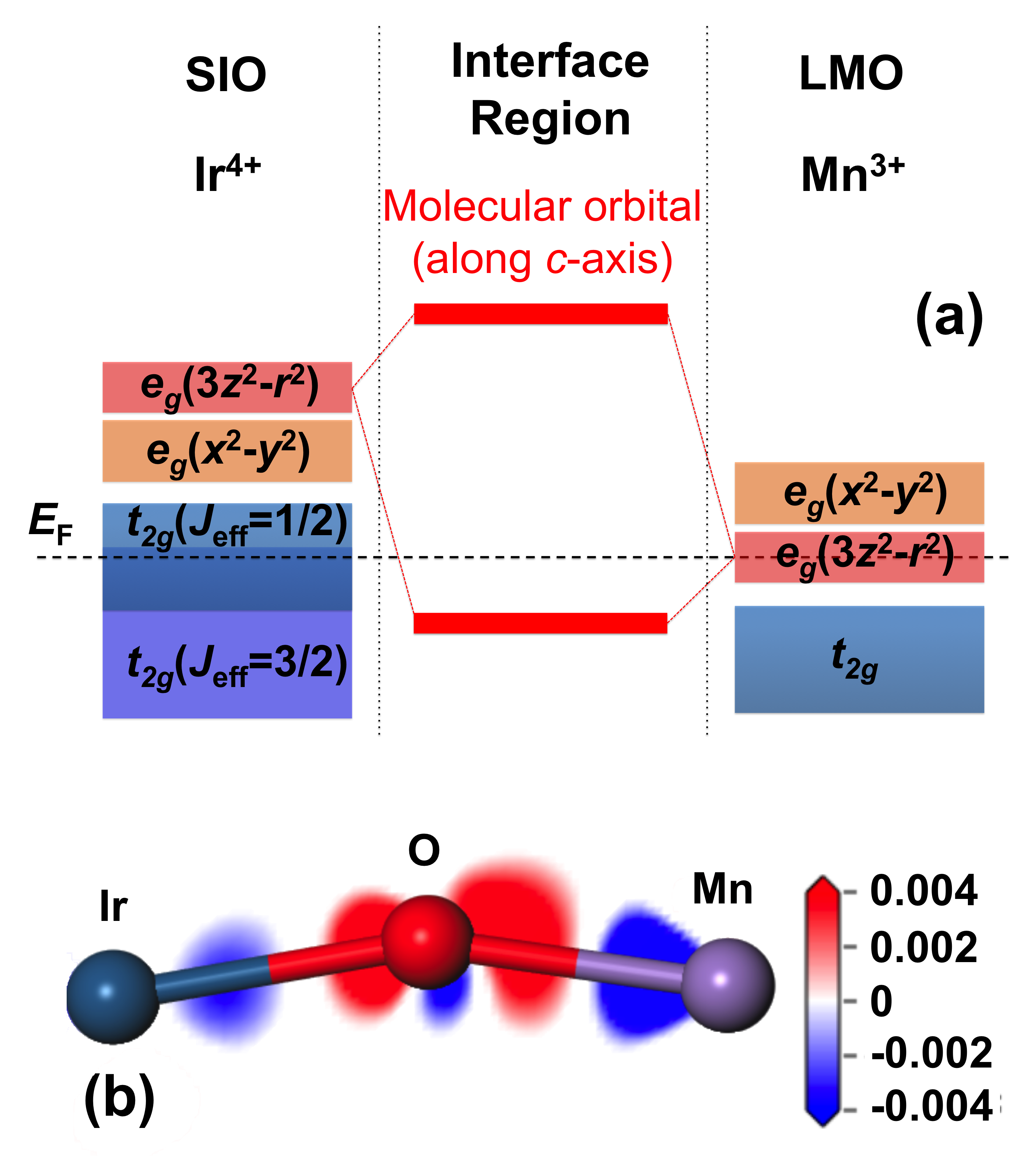}
	\caption{(a) Schematic orbital energy level at the LMO/SIO interface. (b) The total charge density difference among interfacial Mn (purple), O (red), and Ir (navy blue) atoms calculated by DFT. The red and blue distributions denote the electron accumulation and depletion regions, respectively. The scale shows the number of electrons per volume in the unit of \AA$^{-3}$.}
	\label{fig4}
\end{figure}

\begin{figure}
	\includegraphics[width=\columnwidth]{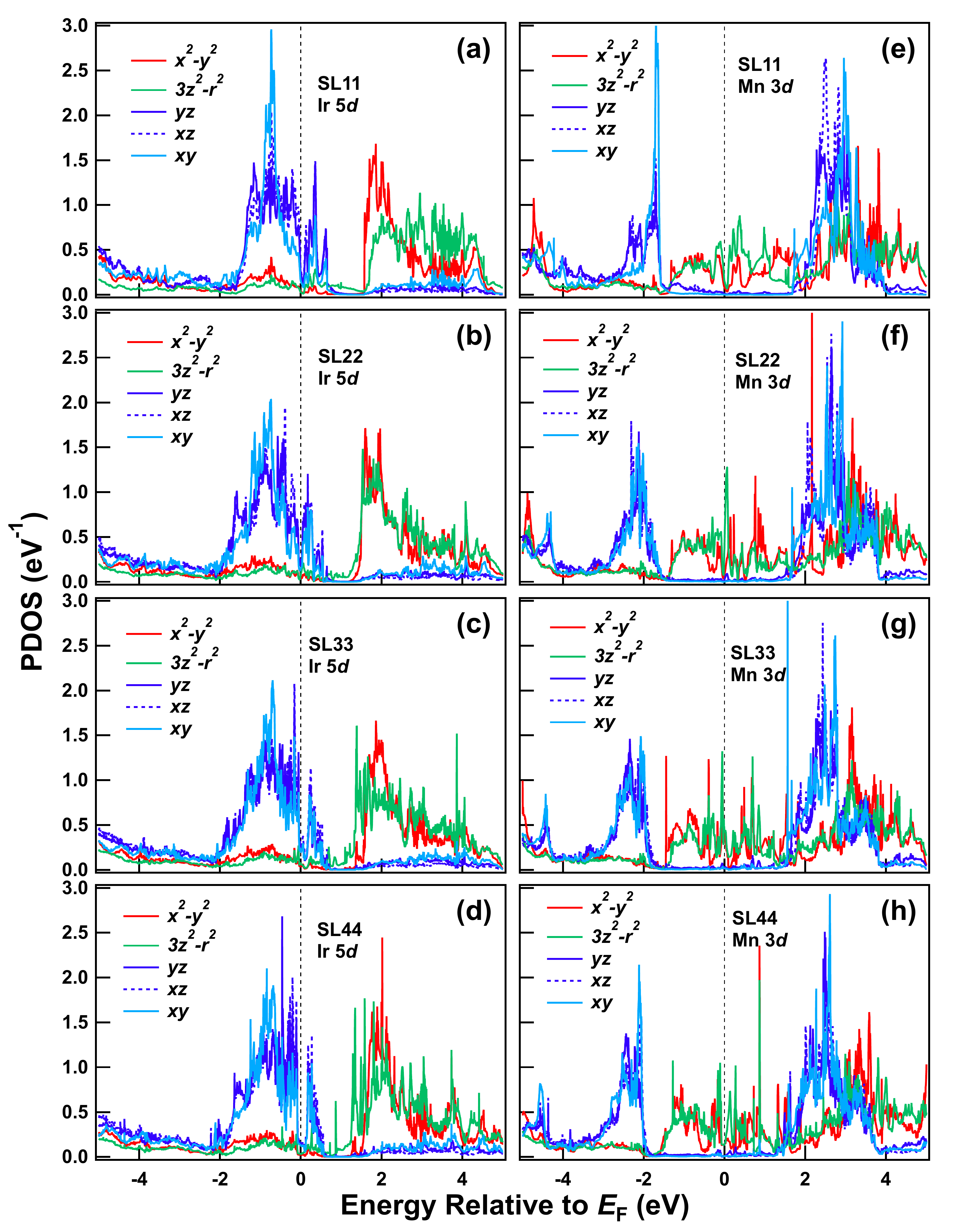}
	\caption{(a-d) Ir 5$d$ and (e-h) Mn 3$d$ PDOS of LMO/SIO SLs calculated by DFT.}
	\label{fig5}
\end{figure}

To understand this SL-period dependent behaviors of the Ir magnetic moments, the electronic structure of the SLs should be investigated in detail. Fig.~\ref{fig4}(a) shows schematic orbital energy level of the LMO/SIO SL system~\cite{27_Okamoto,46_Okamoto}. The octahedral crystal field splits both the Ir 5$d$ and Mn 3$d$ levels into $e_g$ and $t_{2g}$ states. According to the relaxed crystal structures in our DFT calculations, the O-Ir-O bond is compressed while O-Mn-O bond is elongated along the $c$-axis in the SLs. Therefore in SIO layer the $3z^2-r^2$ orbital lies above the $x^2-y^2$ orbital and vice versa in LMO layer due to the Jahn-Teller effect. The $3z^2-r^2$ orbitals of Ir and Mn can hybridize with each other along the $c$-axis and form molecular orbitals~\cite{25_Kim,27_Okamoto}. 
The formation of the molecular orbitals can be visualized by the calculated charge density difference in Fig.~\ref{fig4}(b). Electrons are spatially redistributed from Ir/Mn atoms to the interfacial region near the O atoms. 
The physical picture of the interfacial molecular orbital can also be evidenced by the partial density of states (PDOS), as shown in Fig.~\ref{fig5}. With the decrease of SL period, the $3z^2-r^2$ orbital (mainly located above 1~eV) of Ir obviously shifts to higher energy above the Fermi level ($E_F$), while the $3z^2-r^2$ orbital of Mn exhibits a trend of PDOS redistribution from higher to lower energy.
The SL-period dependence of PDOS is consistent with the formation of molecular orbital that anti-bonding molecular orbital is mainly contributed by Ir and lies at higher energy than the original $3z^2-r^2$ orbital of Ir, while bonding molecular orbital is mainly contributed by Mn and lies at lower energy than the original $3z^2-r^2$ orbital of Mn, as schematically shown in Fig.~\ref{fig4}(a).

\begin{figure}
	\includegraphics[width=\columnwidth]{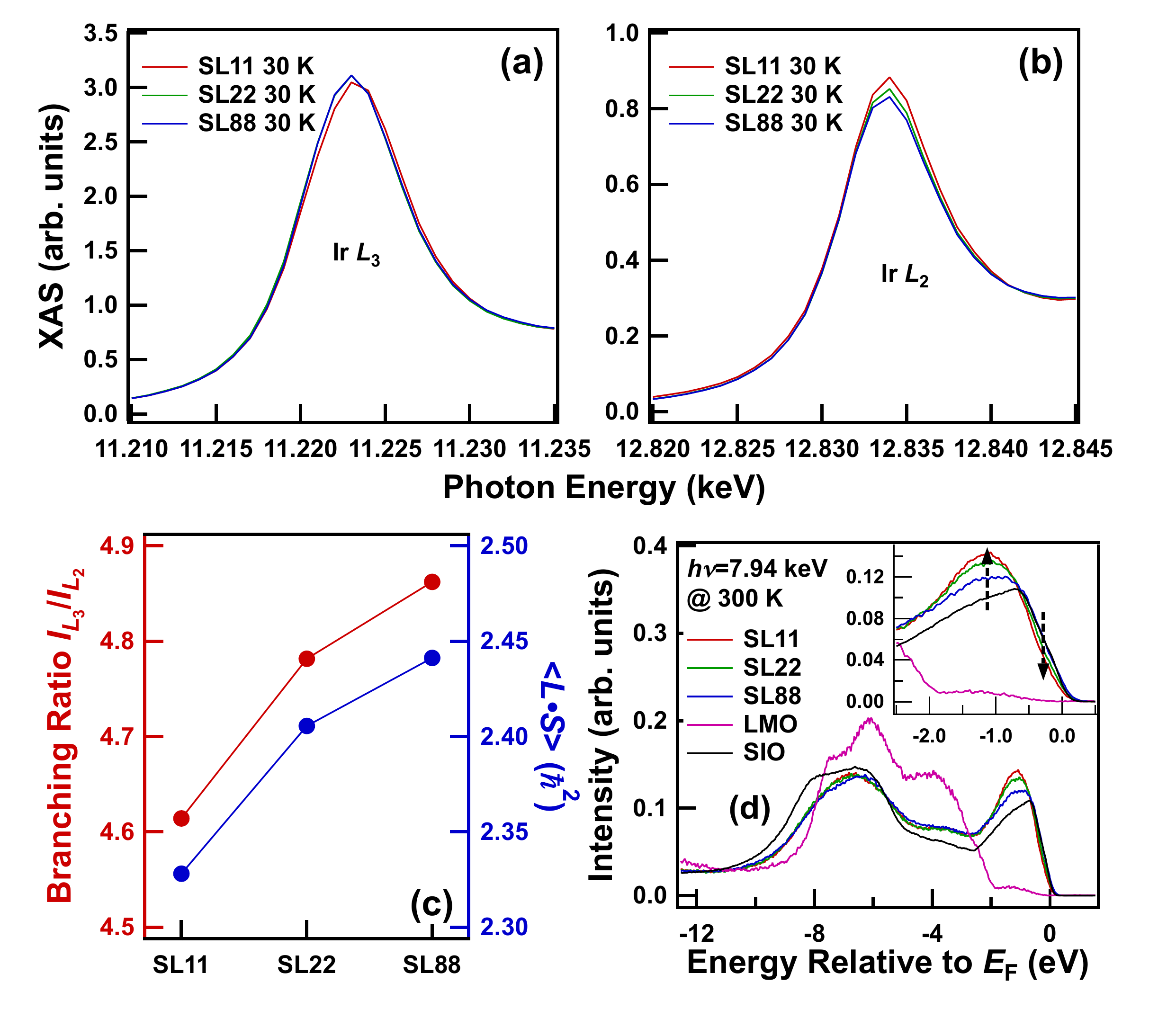}
	\caption{(a,b) Expanded Ir $L$ edge XAS of the LMO/SIO SLs. (c) $BR$ and $\left<L\,{\cdot}\,S\right>$ values of the LMO/SIO SLs. (d) Valence band HAXPES results of LMO/SIO SLs. The spectra are normalized by the total intensity in the binding energy range depicted in the figure. Inset shows the expanded spectra near the $E_F$.}
	\label{fig6}
\end{figure}

Experimental results also show some clues to understand the SL-period dependent electronic structure of the SLs. Details in XAS at the Ir $L$ edge can provide information about the unoccupied Ir 5$d$ states. By taking a closer look at the white-line regions of the XAS (Fig.~\ref{fig6}(a,b)), the white-line intensity at the $L_2$ edge ($I_{L_2}$) obviously increases with the decrease of the SL period, while the white-line intensity of the $L_3$ edge ($I_{L_3}$) keeps nearly constant and slightly decreases for SL11.
This variation of the white-line intensity induces the systematic change of the branching ratio ($BR$) and the expectation value of SOC operator $\left<L\,{\cdot}\,S\right>$ (Fig.~\ref{fig6}(c)). Here $BR=I_{L_{3}}/I_{L_{2}}=(2+r)/(1-r)$,  $r=\left<L\,{\cdot}\,S\right>/n_{h}$~\cite{47_Laan} and $n_{h}=5$. It can be observed that the $BR$ of SL11 is 4.61, which is similar to that of previous reported (LSMO)$_1$/(SIO)$_1$ SL ($BR \approx 4.4$)~\cite{25_Kim}. The $BR$ and $\left<L\,{\cdot}\,S\right>$ value systematically increase with the SL period. Since the $L_2$ edge corresponds to the electric dipole transition from 2$p_{1/2}$ to 5$d_{3/2}$ states while the $L_3$ edge corresponds to the electric dipole transition from 2$p_{3/2}$ to 5$d_{5/2}$ states, the decrease of $BR$ indicates less occupation of 5$d_{3/2}$ states and/or more occupation of 5$d_{5/2}$ states. In perovskite iridates, the octahedral crystal field splits the 5$d_{5/2}$ ($J_{5/2}$) states into $e_g$ states and $J$\textsubscript{eff}=1/2 states, and $J$\textsubscript{eff}=3/2 states originate from the atomic 5$d_{3/2}$ ($J_{3/2}$) states~\cite{11_Kim,44_LagunaMarco}. The change of $BR$ indicates that the occupation of $J$\textsubscript{eff}=3/2 states is decreased by the interfacial coupling and the occupation of $e_g$ states or $J$\textsubscript{eff}=1/2 increases when the interfacial coupling is present. 

We conducted bulk-sensitive HAXPES measurements to further characterize the valence band structure of these SLs (Fig.~\ref{fig6}(d)). 
It can be clearly observed in the inset of Fig.~\ref{fig6}(d) that a feature at a binding energy of $\sim$1.2~eV is enhanced with the decrease of the SL period, which should be the interfacial-coupling-enhanced occupation of the bonding molecular orbital. Simultaneously, the intensity near the $E_F$ decreases with the decrease of the SL period, showing clear evidence that some density of states (DOS) near the $E_F$ was transferred to deeper levels due to the Ir-Mn interfacial coupling.

Enhanced occupation of the $e_g$ states rather than $J$\textsubscript{eff}=3/2 states induced by the interfacial coupling is more likely due to the following reasons. As displayed in the schematic in Fig.~\ref{fig4}(a), the bonding molecular orbital appears below the $E_F$ and changes the relative occupation of different orbitals.  In maganite/iridate SL systems, rather than an ideal $J$\textsubscript{eff}=1/2 scheme, mixed occupation of both $J$\textsubscript{eff}=1/2 and $J$\textsubscript{eff}=3/2 states can often occur~\cite{23_Yi}. When the molecular orbital is formed by interfacial coupling, some of the $J$\textsubscript{eff}=1/2 and $J$\textsubscript{eff}=3/2 electrons of Ir near the $E_F$ will be transferred into the bonding molecular orbital. In particular, the electron transfer from $J$\textsubscript{eff}=3/2 states to the molecular orbital will lead to the change of $BR$ and consequently the local property change of the Ir moments. One may argue that when the SL period increases, the change of Ir 5$d$ bandwidth induced by the dimensionality of SIO layers may also account for the electron redistribution among $J$\textsubscript{eff}=3/2 and $J$\textsubscript{eff}=1/2 states and give rise to the change of $BR$. However, as the SL period increases, the enhanced Ir-Ir hopping will result in simultaneous increase of the bandwidth of both $J$\textsubscript{eff}=3/2 and $J$\textsubscript{eff}=1/2 states. Since $J$\textsubscript{eff}=3/2 states are nearly fully occupied, the center of $J$\textsubscript{eff}=3/2 states lies deep below the $E_F$. While the center of $J$\textsubscript{eff}=1/2 states lies close to the $E_F$, as schematically shown in Fig.~\ref{fig4}(a). With the center of mass of the states fixed, widening of the $J$\textsubscript{eff}=3/2 states should lead to more DOS above $E_F$ and less occupation of itself, which does not agree with the change of XAS white-line intensity. Consequently, the experimentally observed $BR$ change should be mainly attributed to the electron transfer between 
$J$\textsubscript{eff}=3/2 states and the $e_g$ states, but not simply the redistribution of electrons within the $t_{2g}$ ($J$\textsubscript{eff}=3/2 and $J$\textsubscript{eff}=1/2) states. 

Up to now, we illustrated the effect of interfacial Ir-Mn coupling on the electronic structure in LMO/SIO SL system. Based on this, we can discuss the consequent effect on the $m_o$/$m_{s(e)}$ ratio. Since $J$\textsubscript{eff}=3/2 states originate from the atomic 5$d_{3/2}$ ($J_{3/2}$) states~\cite{11_Kim}, whose spin and orbital moments are antiparallel. While $J$\textsubscript{eff}=1/2 and $e_g$ states originate from the atomic 5$d_{5/2}$ ($J_{5/2}$) states~\cite{11_Kim}, whose spin and orbital moments are parallel. When Ir-Mn interfacial coupling transfers some electrons from $J$\textsubscript{eff}=3/2 states to $e_g$ states, the SOC sign of these electrons are effectively changed. Spin and orbital components of the Ir moments become more parallel. As a result, we observed smaller $m_o$/$m_{s(e)}$ ratio in SLs with shorter SL period, in which the interfacial coupling is more efficient. 
From another angle of view, the consistent trend of $m_o$/$m_{s}$ ratio in DFT calculations further reveals that $m_s$ is more sensitive to the interfacial coupling than $m_o$. This is due to the fact that spin is directly carried by the redistributed electrons from $t_{2g}$ states to $e_g$ states. The shorter SL period, the more electrons transferred from Ir $t_{2g}$ to $3z^2-r^2$ molecular orbitals, and the more Ir $t_{2g}$ spins which are originally antiparallel to the total Ir spin tend to reverse its direction and align parallel to the total Ir spin due to the Hund's coupling~\cite{27_Okamoto}.
On the other hand, the $m_o$ is relatively robust to the interfacial coupling, which drives the $m_o$/$m_{s}$ ratio to be smaller when the SL period decreases.

\section*{Conclusions}
In conclusion, by comprehensive experimental and theoretical investigations of the LMO/SIO SL system, we conclude that the local properties of Ir FM moments and interfacial electronic structure can be modified by the SL period. Hybridization between the $3z^2-r^2$ orbitals of Ir and Mn along the $c$-axis of the SLs can form a bonding molecular orbital which lies below the $E_F$, so that electrons from Ir $t_{2g}$ states near the $E_F$ are pulled down into this bonding molecular orbital. 
The $m_o$/$m_{s(e)}$ ratio of Ir can be modified by the interfacial coupling due to the electron transfer between Ir $t_{2g}$ and $e_g$ orbitals, which have different spin alignments. Our results demonstrate a clear physical picture of the Ir-Mn interfacial coupling in manganite-iridate SL system. The conclusions could also be generalized to other TMO-based perovskite heterostructures and SL systems, such as SrRuO$_3$/SrIrO$_3$ heterostructures~\cite{18_Matsuno}, etc.

\section*{Acknowledgments}
This work was supported by Grant-in-Aid for JSPS fellows (No. 17F17327) and JSPS KAKENHI (Grant No. 19H05824), Japan, as well as the Natural Science Foundation of China (Grant No. 51729201) . The synchrotron radiation experiments at SPring-8 were performed under the approval of the Japan Synchrotron Radiation Research Institute (Proposal No. 2018A1232, 2018B1449 and 2019A1239). X.~R.~Wang acknowledges supports from the Nanyang Assistant Professorship grant from Nanyang Technological University and Academic Research Fund Tier 1 (RG108/17 and RG177/18) from Singapore Ministry of Education. L.~Shen acknowledges the support from Singapore MOE Tier 1 (Grant R-265-000-615-114). We also acknowledge the enlightening discussion with Prof. M.~van~Veenendaal, as well as the experimental supports and discussions provided by Dr. K.~Ishii and Dr. Y.~Takeda.

\section*{Appendix A}
\begin{figure}
	\includegraphics[width=\columnwidth]{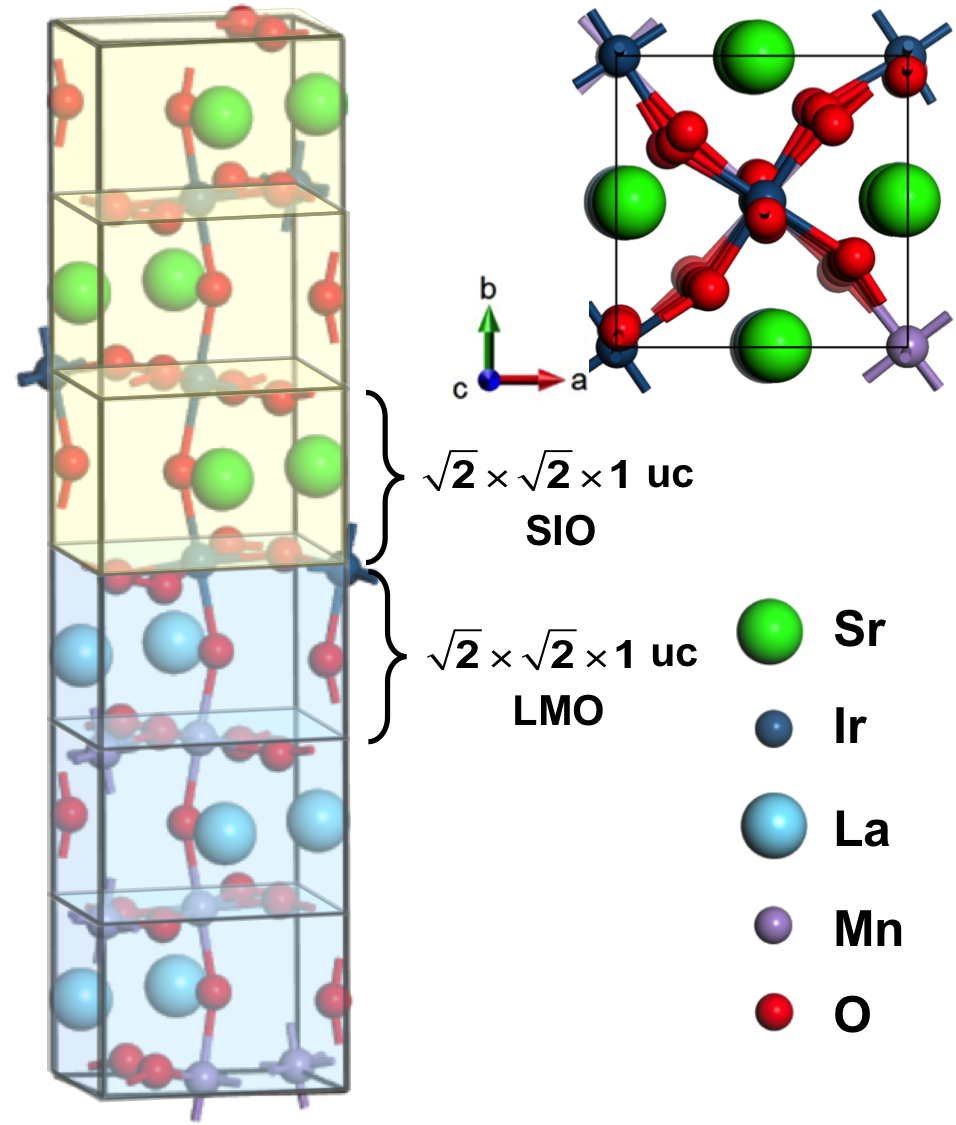}
	\caption{The basic doubled unit cell used for the DFT calculations and the relaxed crystal structure of the SL33 which comprises 6 perovskite layers in the (001) orientation with 60 atoms in total.}
	\label{fig7}
\end{figure}

Fig.~\ref{fig7} shows the detailed lattice structure used for the DFT calculations. SL33 is taken as an example. Our theoretical approaches have been validated by comparing with previously reported experimental benchmarks~\cite{27_Okamoto,48_Elemans,49_Lee}.

\section*{Appendix B}
\begin{figure}
	\includegraphics[width=\columnwidth]{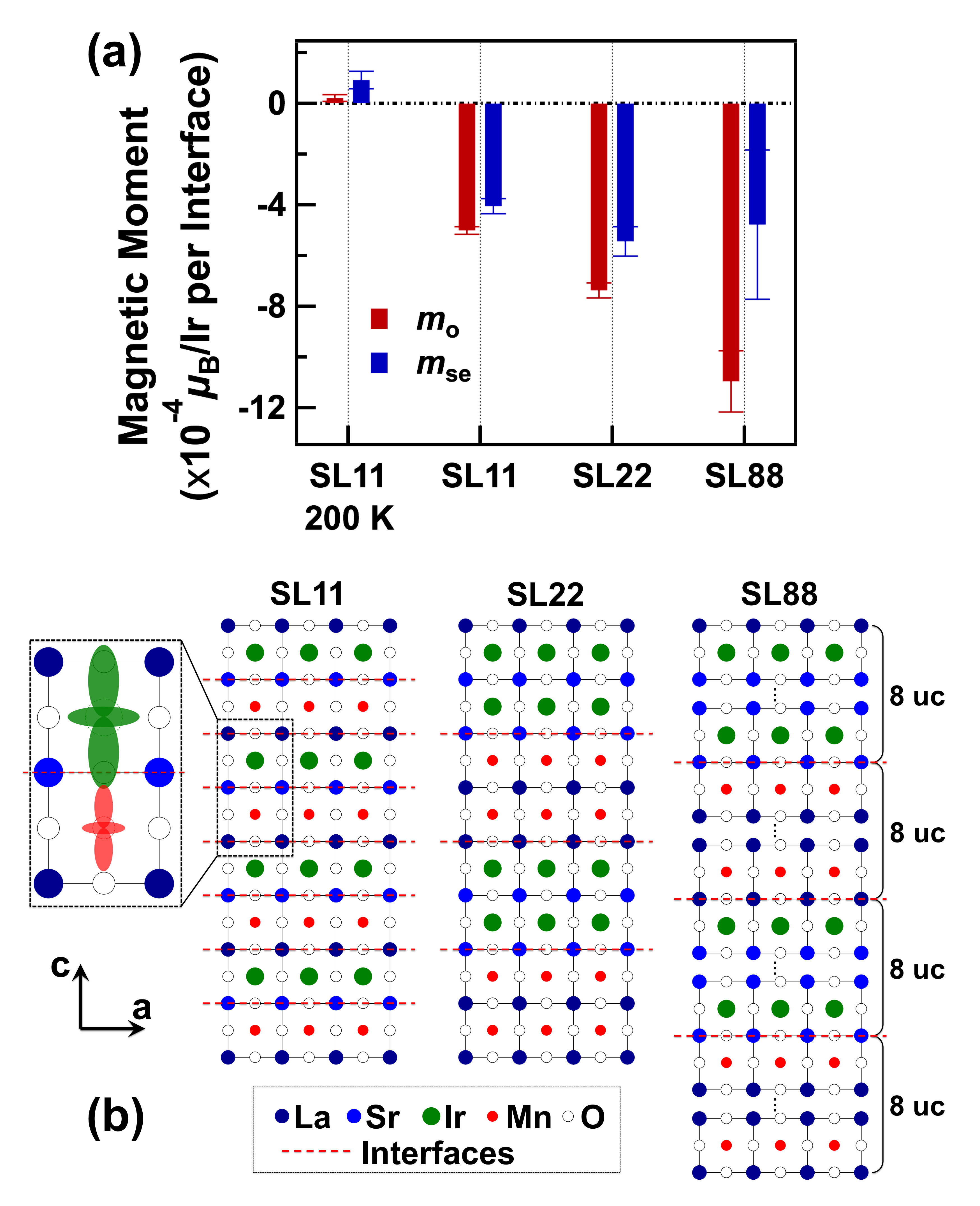}
	\caption{(a) Ir magnetic moments averaged to each interface. (b) Crystal structure of the LMO/SIO SLs. The interfacial hybridization geometry of $3z^2-r^2$ orbitals of Ir and Mn is depicted in the enlarged figure.}
	\label{fig8}
\end{figure}

To get a feeling about the spatial range that the interfacial coupling can influence, we averaged the magnetic moments obtained by sum-rules analysis to each interface. As shown in Fig.~\ref{fig8}(a), the size of Ir magnetic moment per interface increases with the SL period, which indicates that the effect of the interfacial coupling may not be restricted only in the unit cells adjacent to the interface, especially by comparing SL22 and SL88. On the other hand, the difference between SL11 and SL22/SL88 can be attributed to other factors. As depicted in Fig.~\ref{fig8}(b), in SL11 every SIO unit cell is sandwiched by two LMO unit cells, and in SL22 and SL88, every SIO unit cell is only adjacent to one Ir-Mn interface. So in SL11, every Ir $3z^2-r^2$ orbital is directly hybridizing with two Mn $3z^2-r^2$ orbitals simultaneously. While in SL22 and SL88, every Ir $3z^2-r^2$ orbital has only one counterpart for interfacial hybridization. This factor may cause different electronic structure and local properties of Ir moments at the interface.

\bibliographystyle{unsrt}

\end{document}